\documentclass[12pt,preprint]{aastex}

\newcommand{\kms}{km~s$^{-1}$}
\newcommand{\etal}{{\it et al.\/}}
\newcommand{\subsun}{\mbox{$_{\odot}$}}

\begin{document}

\title{The Not So Extraordinary Globular Cluster M31~037--B327\altaffilmark{1}}

\author{Judith G. Cohen\altaffilmark{2}}

\altaffiltext{1}{Based on observations obtained at the
W.M. Keck Observatory, which is operated jointly by the California 
Institute of Technology, the University of California, and the
National Aeronautics and Space Administration.}

\altaffiltext{2}{Palomar Observatory, Mail Stop 105-24,
California Institute of Technology, Pasadena, Ca., 91125}

\begin{abstract}

A velocity dispersion has been measured for the luminous
globular cluster M31~037--B327, claimed to be the
most massive star cluster in the Local Group and to
be a young ``super star cluster'' that has survived
to an old age.  M31~037--B327
has a mass comparable to that of M31~G1, but not
significantly larger.  Although near the upper end for
the mass distribution of globular clusters, it is not an 
unprecedented extraordinary object.

\end{abstract}

\keywords{galaxies: individual (M31), galaxies: star clusters}

\section{Introduction}

The globular cluster 037--B327 is an extremely red non-stellar
object close to the disk of M31.  \cite{barmby02a} calculated
its reddening, using photometry in the compiled catalog
of M31 globular clusters of \cite{barmby00}, to be $E(B-V) = 1.30\pm0.04$~mag,
and from this inferred that it was extremely
luminous, with $M_V = -11.7$~mag, making it
the most luminous globular cluster in M31.  The recent compilation
of data for M31 globular clusters by the Bologna group 
\citep{galletti04} also confirms the unusual properties
of this object. 

Recently \cite{ma06a} have studied this cluster in more detail.
They compare their new multicolor photometry to  theoretical
spectral energy distributions of varying ages to determine
the reddening and age of this cluster.  They find
that it has a photometric mass of $3.0\pm0.5 \times 10^7 M$\subsun;
they then claim this object to be the most massive star cluster of any age
in the Local Group.  This claim is based in part on data from
the compiled catalogs of \cite{barmby00} for the M31 system and of
\cite{harris96} for the Milky Way globular
cluster system. \cite{ma06a}  predict that M31~037--B327
has a 1D velocity dispersion of $72\pm13$~\kms,~  far larger
than that of any other known M31 globular cluster.  Velocity dispersions
for a number of the brightest globular clusters in M31
were determined by \cite{george}; the highest
value they measured was $25.1\pm 0.3$~\kms, for M31 G1.

\section{$\sigma_v$ for M31~037--B327}

We have obtained high spectral resolution spectra with
HIRES at Keck \citep{vogt94} of M31~037--B327
and of M31~G1, as well as of several metal-rich giants
on the upper RGB of the Draco dwarf spheroidal galaxy
which served as template stars.    We took two 1500 sec
exposures of the object of interest, one 600 sec exposure of the much
brighter (at $V$) object  M31~G1, and several suitable spectra of
template red giants from the same run in early September 2006.
All these spectra were taken with a $1.1^{\arcsec} \times 7^{\arcsec}$ 
slit, giving a spectral resolution of 35,000 with 5 pixels/spectral
resolution element ($\sim$1.3~km sec$^{-1}$ pixel$^{-1}$).

\begin{figure}
\epsscale{0.8}
\plotone{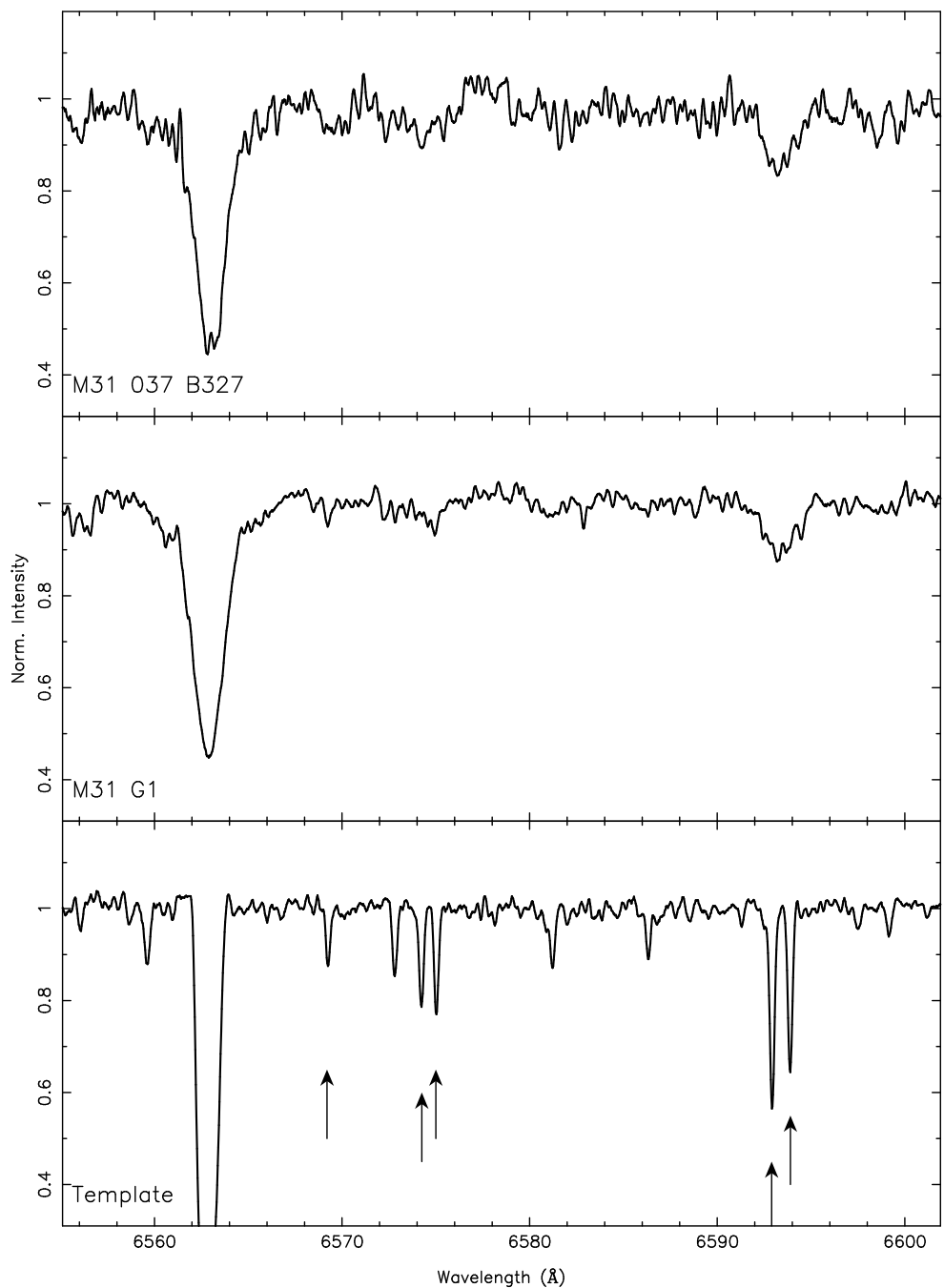}
\caption{The region of the spectrum including H$\alpha$ 
is shown for M31~037--B327 (top panel), M31~G1 (middle panel),
and a template luminous red giant.  The spectra have been 
normalized, shifted to the rest frame, and slightly smoothed
by a Gaussian with FWHM = 5~\kms.
The spectral features marked by arrows in the bottom panel are 
all Fe~I lines.
\label{figure_vsigma}}
\end{figure}

We chose for analysis 
parts of three different echelle orders
with strong broadened spectral features visible in the 
spectrum of  M31~037--B327.  Since the flux
for this object was dropping rapidly towards the blue,
but the number of strong lines was dropping rapidly
towards the red, this was something of a compromise.
Cutouts of the spectra in one of the regions analyzed, 
that near H$\alpha$, are shown for 
the two M31 globular clusters and one of the
template objects in Fig.~\ref{figure_vsigma}.  It is immediately
apparent that $\sigma_v$ for M31~037--B327 is comparable
to that of M31~G1, but is not significantly larger.
This impression is sustained when one examines the strongest
lines in these spectra over their full useful wavelength range, 
which for the object of interest is from about 5100 to 8350~\AA.

We used the Fourier
quotient method of \cite{sargent} to determine $\sigma_v$.  
We subsequently applied
an aperture correction factor of 1.14 following \cite{george}.
The resulting $\sigma_v$ are given in Table~\ref{table_vsigma}.
Our measured $\sigma_v$ for M31~G1 is in good agreement with that of
\cite{george}, $25.1\pm0.3$~\kms.

\section{Comparison of M31~037--B327 With M31~G1}

M31~G1, if one ignores M31~037--B327, is widely believed to 
be the most luminous globular cluster
in M31, see, e.g. \cite{meylan01}. \cite{gebhardt} suggested,
based on HST/STIS spectroscopy, that it contains
a 20,000~$M$\subsun\ central black hole.
\cite{baumgardt} dispute this; they obtain a good fit to
all available data for G1 with their dynamical model, which does
not include an intermediate-mass central black hole.

\cite{ma06b} utilize an ACS image from HST to determine
the structural parameters of M31~037--B327, and discuss the possibility,
suggested earlier by \cite{barmby02a},
that this object is the nucleus of a dwarf galaxy accreted by M31.
The ACS image reveals a dust lane crossing the face
of the cluster.  Toy models of a uniform extinction over
most of the cluster with much larger $E(B-V)$ over a
smaller part of the cluster suggest substantial errors
in the values of the dereddened fluxes 
can occur in such circumstances if a standard
extinction curve as a function of wavelength is applied assuming 
a constant $E(B-V)$.
The reddening of G1 is small as it has a projected distance from the center
of M31 of $\sim$40 kpc; we assume only the foreground
Galactic reddening applies.

We compare the properties of M31~037--B327 with those of G1.
The values of $\sigma_v$ are comparable (see Table~\ref{table_vsigma}), 
with that of G1 perhaps being slightly larger.
Since the reddening of M31~037--B327 is large and spatially
variable across the face of the cluster, we prefer to
compare the luminosities at $K$ where the impact of the reddening
is minimized. We assume both objects are
old stellar clusters.
This comparison is sure to be more reliable
than a similar one making the same assumption
carried out at $V$ by \cite{ma06a} which
suggested that M31~037--B327 is a factor of $\sim$2.5 times
more luminous than G1.  

In  late September 2006 we acquired 
an image at $K$ of the field of M31~037--B327
with the Wide Field Infrared Camera \citep{wilson03} at the Hale Telescope
on Palomar Mountain to verify the IR photometry of this
cluster.  We find $K_s = 11.06\pm0.05$~mag
in the 2MASS \citep{2mass1,2mass2} system for an aperture 
$11^{\arcsec}$ in diameter.  This 
is within 0.01~mag of that
derived from the 2MASS database
by \cite{galletti04}\footnote{Table~2 of \cite{galletti04} presents
IR mags measured from 2MASS transformed into the CIT
system using the transformation equation of 
\cite{carpenter01}.  We made the required inverse transformation 
for the comparison quoted above.}.
$K_s$ for G1
from 2MASS as reported by
\cite{galletti04} is 0.025 mag fainter than that
of  M31~037--B327.
If one takes $E(B-V)$ for the heavily reddened 
cluster M31~037--B327
as $1.3$~mag with a more realistic error than that adopted by
\cite{ma06a},
given the spatially varying reddening, of $\pm 0.3$~mag,
then the M31~037--B327 has $M_K$ brighter by 0.16$\pm 0.03$ mag.
We thus find that M31~037--B327 has 
$L_K$ comparable to that of G1 and at most 20\% larger, even
allowing for a generous
uncertainty in $E(B-V)$.
$M_K$ is well known to be a good measure of the 
total luminosity for old stellar
systems; the dependence of $L_{\lambda}$ on [Fe/H] is smaller at 2.4$\mu$
than at optical wavelengths. 

In order to determine the virial mass of each of these objects,
we must assume that the same initial mass function prevails
in both of these clusters, and combine 
$\sigma_v$ with the half light radius determined from HST
or other high spatial resolution imaging.  \cite{ma06b} have
measured $R_h$ for M31~037--B327, while there are two independent
and discrepant determinations of $R_h$ for G1 (that of Meylen \etal\ 2001
and of Barmby, Holland \& Huchra 2002).  $R_h$ for the highly reddened cluster
M31~037--B327 is not larger than that of G1.

Thus there is no evidence that the highly reddened globular cluster
M31~037--B327 is substantially more massive than M31~G1.

\section{Conclusions}

The luminous object  M31~037--B327 believed to be a globular
cluster in M31 is indeed a massive
object which may or may not be
the nucleus of an accreted galaxy, as was suggested by \cite{meylan01} and others.  
However,
although it is among the most massive globular
clusters in M31, at least four other globular clusters, G1, G78, G280 and G213,
studied by \cite{george}, whose earlier $\sigma_v$ for G1 is
confirmed here,
have comparable masses. The first three of these
are probably more massive than M31~037--B327. Among the Galactic globular
clusters, $\omega$ Cen and NGC~6441 have $\sigma_v$ comparable
to that of M31~037--B327, based on the compilation
of \cite{pryor}.   Contrary to the conclusion of \cite{ma06a}, we find
it to be very similar to the well studied cluster M31 G1.
M31~037--B327 is not an extraordinarily massive old cluster, and
is almost certainly not the most massive
cluster of any age in the Local Group.

\acknowledgements

The entire Keck/HIRES user communities owes a huge debt to 
Jerry Nelson, Gerry Smith, Steve Vogt, and many other 
people who have worked to make the Keck Telescope and HIRES  
a reality and to operate and maintain the Keck Observatory. 
We are grateful to the W. M.  Keck Foundation for the vision to fund
the construction of the W. M. Keck Observatory. 
The authors wish to extend special thanks to those of Hawaiian ancestry
on whose sacred mountain we are privileged to be guests. 
Without their generous hospitality, none of the observations presented
herein would have been possible.
This publication makes use of data from the Two Micron All-Sky Survey,
which is a joint project of the University of Massachusetts and the 
Infrared Processing and Analysis Center, funded by the 
National Aeronautics and Space Administration and the
National Science Foundation.
J.G.C. is grateful to NSF grant AST-0507219  for partial support.


\clearpage

\begin{deluxetable}{l rrrr}
\tablenum{1}
\tablewidth{0pt}
\small
\tablecaption{Velocity Dispersion Measurements for 2 M31 Globular Clusters
\label{table_vsigma}}
\tablehead{ \colhead{ID} & \colhead{Wavelength Range}  &
\colhead{$\sigma_v$}   & \colhead{Uncertainty} 
&  \colhead{Ap.Cor. $\sigma_v$\tablenotemark{a}}\\
\colhead{} & \colhead{(\AA)} &  \colhead{(1$\sigma$, \kms)} &
\colhead{(\kms)} & \colhead{(\kms)}
}
\startdata
M31~037--B327 & 5150--5190 & 19.2 & 3.5 & 21.9 \\
~~~ & 6105--6190 & \nodata\tablenotemark{b} & \nodata & \nodata \\
~             & 6545--6600 & 19.9 & 3.4 & 22.6 \\
~~ \\
M31~G1 &  5150--5190 & 19.6 & 1.8 & 22.4 \\
~~~ & 6105--6190 & 22.3 & 1.5 & 25.4 \\
~~~ &    6545--6600 & 22.4 & 1.5 & 25.6 \\
\enddata
\tablenotetext{a}{Aperture correction set to a factor of 1.14
following \cite{george}.}
\tablenotetext{b}{Features were too weak to permit a reliable
determination of $\sigma_v$ in this spectral region.}
\end{deluxetable}

\clearpage

\end{document}